\documentclass{acm_proc_article-sp}
\usepackage{subfigure}
\makeatletter
\newif\if@restonecol
\makeatother

\usepackage[ruled]{algorithm2e} 
\begin{document}

\title{Influence and Passivity in Social Media}


\numberofauthors{4} %  in this sample file, there are a *total*
\author{
% 1st. author
\alignauthor
Daniel M. Romero\\
       \affaddr{Cornell University}\\
       \affaddr{Center for Applied Mathematics}\\
       \affaddr{Ithaca, New York, USA}\\
       \email{dmr239@cornell.edu}\\ 
% 2nd. author
\alignauthor
Wojciech Galuba\\
       \affaddr{EPFL}\\
        \affaddr{Distributed Information Systems Lab}\\
         \affaddr{Lausanne, Switzerland}\\
       \email{wojciech.galuba@epfl.ch}\\ 
% 3rd. author
\and
\alignauthor 
Sitaram Asur\\
	\affaddr{Social Computing Lab}\\
       \affaddr{HP Labs}\\
       \affaddr{Palo Alto, California, USA}\\
       \email{sitaram.asur@hp.com}
  % use '\and' if you need 'another row' of author names
% 4th. author
\alignauthor 
Bernardo A. Huberman\\
      \affaddr{Social Computing Lab}\\
	\affaddr{HP Labs}\\
      \affaddr{Palo Alto, California, USA}\\
      \email{bernardo.huberman@hp.com}
}
% There's nothing stopping you putting the seventh, eighth, etc.
% author on the opening page (as the 'third row') but we ask,
% for aesthetic reasons that you place these 'additional authors'
% in the \additional authors block, viz.
%\additionalauthors{Additional authors: John Smith (The Th{\o}rv{\"a}ld Group,
%email: {\texttt{jsmith@affiliation.org}}) and Julius P.~Kumquat
%(The Kumquat Consortium, email: {\texttt{jpkumquat@consortium.net}}).}
% Just remember to make sure that the TOTAL number of authors
% is the number that will appear on the first page PLUS the
% number that will appear in the \additionalauthors section.

\maketitle
\begin{abstract}
\label{abstract}
The ever-increasing amount of information flowing through
Social Media forces the members of these networks to compete for attention and influence by relying on other people to spread their message.
A large study of information propagation within Twitter reveals
that the majority of users act as passive information
consumers and do not forward the content to the network.
Therefore, in order for individuals to become influential they
must not only obtain attention and thus be popular, but also
overcome user passivity. We propose an algorithm that determines
the influence and passivity of users based on
their information forwarding activity. An evaluation performed
with a 2.5 million user dataset shows that our influence measure
is a good predictor of URL clicks, outperforming several
other measures that do not explicitly take user passivity into
account. We also explicitly demonstrate
that high popularity does not necessarily imply high
influence and vice-versa.
\end{abstract}

%\category{H.4}{Categories}{Subject Description}
%A category including the fourth, optional field follows...

%\terms{Terms}

\section{Introduction} 
\label{sec-introduction}
The explosive growth of Social Media  has provided millions of people the
opportunity to create and share content on a scale barely imaginable a few years
ago. Massive participation in these social networks is reflected  in the
countless number of opinions, news and product reviews that are constantly posted
and discussed in social sites such as Facebook, Digg and Twitter, to name
a few. Given this widespread generation and consumption of content, it is natural
to target one's messages to highly connected people who will propagate
them further in the social network. This is particularly the case
in Twitter, which is one of the fastest growing  social networks on the
Internet, and thus the focus of advertising companies and celebrities eager to exploit this
vast new medium. As a result, ideas, opinions, and products compete with all
other content for the scarce attention of the user community. In spite of the
seemingly chaotic fashion with which all these interactions take place, certain
topics manage to get an inordinate amount of attention, thus bubbling to the top
in terms of popularity and contributing to new trends and to the public agenda of
the community. How this happens in a world where crowdsourcing dominates is still
an unresolved problem, but there is considerable consensus on the fact that two
aspects of information transmission seem to be important in determining which
content receives inordinate amounts  of attention.

One is the popularity and status of given members of these social networks, which
is measured by the level of attention they receive in the form of followers who
create links to their accounts to automatically receive the content they
generate. The other is the influence that these individuals wield, which is
determined by the actual propagation of their content through the network. This
influence is determined by many factors, such as the novelty and resonance of
their messages with those of their followers and the quality and frequency of the
content they generate. Equally important is the passivity of members of the network
which provides a barrier to propagation that is often hard to overcome. Thus gaining knowledge
of the identity of influential and least passive people in a network can be
extremely useful from the perspectives of viral marketing, propagating one's point of view, as well as
setting which topics dominate the public agenda.

In this paper, we analyze the propagation of web links on twitter over time to
understand how attention to given users and their influence is determined. We
devise a general model for influence using the concept of passivity in a social
network and develop an efficient algorithm similar to the HITS
algorithm~\cite{HITS} to quantify the influence of all the users in the network.
Our influence measure utilizes both the structural properties of the network as
well as the diffusion behavior among users. The influence of a user thus depends
on not only the size of the influenced audience, but also on their
passivity. This differentiates it from earlier measures of influence which were primarily
based on individual statistical properties such as the number of followers or
retweets~\cite{Cha2010}.

We have shown through our extensive evaluation that this influence model
outperforms other measures of influence such as PageRank, H-index, the number of
followers and the number of retweets. In addition it has good predictive
properties in that it can forecast in advance the upper bound on the number of
clicks a URL can get. We have also presented case studies showing the top
influential users uncovered by our algorithm. An important conclusion from the results is that the correlation
between popularity and influence is quite weak, with the most influential users
not necessarily the ones with the highest popularity. Additionally, when we
considered nodes with high passivity, we found a majority of them to be spammers
and robot users. This demonstrates an application of our algorithm in
automatic user categorization and filtering of online content .

\section{Related work} \label{sec-related-work}
The study of information and influence propagation in social networks has been
particularly active for a number of years in fields as disparate as sociology,
communication, marketing, political science and physics. Earlier work
focused on the effects that scale-free networks and the affinity of their
members for certain topics had on the propagation of information ~\cite{Wu2004}. Others
discussed the presence of key influentials~\cite{Domingos2001,Goyal2010,Agarwal2008,Weng2010, Watts2007} in a social
network, defined as those who are responsible for the overall information
dissemination in the network.  This research highlighted the value of highly
connected individuals as key elements in the propagation of information through
the network.

Huberman et al.~\cite{Huberman2008} studied the social interactions on Twitter
to reveal that the driving process for usage is a sparse hidden network underlying
the friends and followers, while most of the links represent meaningless
interactions. Jansen et al.~\cite{Jansen2009} have examined twitter as a mechanism for word-of-mouth
advertising. They considered particular brands and products and examined the
structure of the postings and the change in sentiments. Galuba et
al.~\cite{galuba-wosn10} propose a propagation model that predicts, which
users will tweet about which URL based on the history of past user activity.

There have also been earlier studies focused on social influence and propagation.
Agarwal et al. ~\cite{Agarwal2008} have examined the problem of identifying
influential bloggers in the blogosphere. They discovered that the most
influential bloggers were not necessarily the most active. Aral et
al~\cite{Aral2009} have distinguished the effects of homophily from influence as
motivators for propagation. As to the study of influence within twitter, Cha et
al.~\cite{Cha2010} have performed a comparison of three different measures of
influence - indegree, retweets and user mentions. They discovered that while
retweets and mentions correlated well with each other, the indegree of users did
not correlate well with the other two measures. Based on this, they hypothesized
that the number of followers may not a good measure of influence.  On the other
hand, Weng et al~\cite{Weng2010} have proposed a topic-sensitive PageRank measure
for influence in Twitter. Their measure is based on the fact that they observed
high reciprocity among follower relationships in their dataset, which they
attributed to homophily. However, other work~\cite{Cha2010} has shown that the
reciprocity is low overall in Twitter and contradicted the assumptions of this
work.

\section{Twitter} \label{sec-twitter}

\subsection{Background on Twitter}
Twitter is an extremely popular online microblogging service, that has gained a
very large user base, consisting of more than 105 million users (as of April
2010). The Twitter graph is a directed social network, where each user chooses to
follow certain other users. Each user submits periodic status updates, known as
\emph{tweets}, that consist of short messages limited in size to 140 characters.
These updates typically consist of personal information about the users, news or
links to content such as images, video and articles. The posts made by a user are
automatically displayed on the user's profile page, as well as shown to his
followers.

A \emph{retweet} is a post originally made by one user that is forwarded by
another user. Retweets are useful for propagating interesting posts and links
through the Twitter community.

Twitter has attracted lots of attention from corporations for the immense
potential it provides for viral marketing. Due to its huge reach, Twitter is
increasingly used by news organizations to disseminate news updates, which are
then filtered and commented on by the Twitter community. A number of businesses
and organizations are using Twitter or similar micro-blogging services to
advertise products and disseminate information to stockholders.

\subsection{Dataset} \label{sec-twitter-dataset}

Twitter provides a Search API for extracting tweets containing particular
keywords. To obtain the dataset for this study, we continuously queried the
Twitter Search API for a period of 300 hours starting on 10 Sep 2009 for all
tweets containing the string \texttt{http}. This allowed us to acquire a
continuous stream of 22 million tweets with URLs, which we estimated to be 1/15th
of the entire Twitter activity at that time. From each of the accumulated tweets,
we extracted the URL mentions. Each of the unique 15 million URLs in the data set
was then checked for valid formatting and the URLs shortened via the services
such as \texttt{bit.ly} or \texttt{tinyurl.com} were expanded into their original
form by following the HTTP redirects. For each encountered unique user ID, we
queried the Twitter API for metadata about that user and in particular the user's
followers and followees. The end result was a dataset of timestamped URL mentions
together with the complete social graph for the users concerned.

\textbf{User graph.} The user graph contains those users whose tweets
appeared in the stream, i.e., users that during the 300 hour observation
period posted at least one public tweet containing a URL. The graph does not
contain any users who do not mention any URLs in their tweets or users that have
chosen to make their Twitter stream private.

For each newly encountered user ID, the list of followed users was only fetched
once. Our data set does not capture the changes occurring in the user graph over
the observation period.
\section{The IP Algorithm} \label{sec-algorithm}

\begin{figure}
 \label{fig-retweet-rate}
   \centering
     \includegraphics[width=0.46\textwidth]{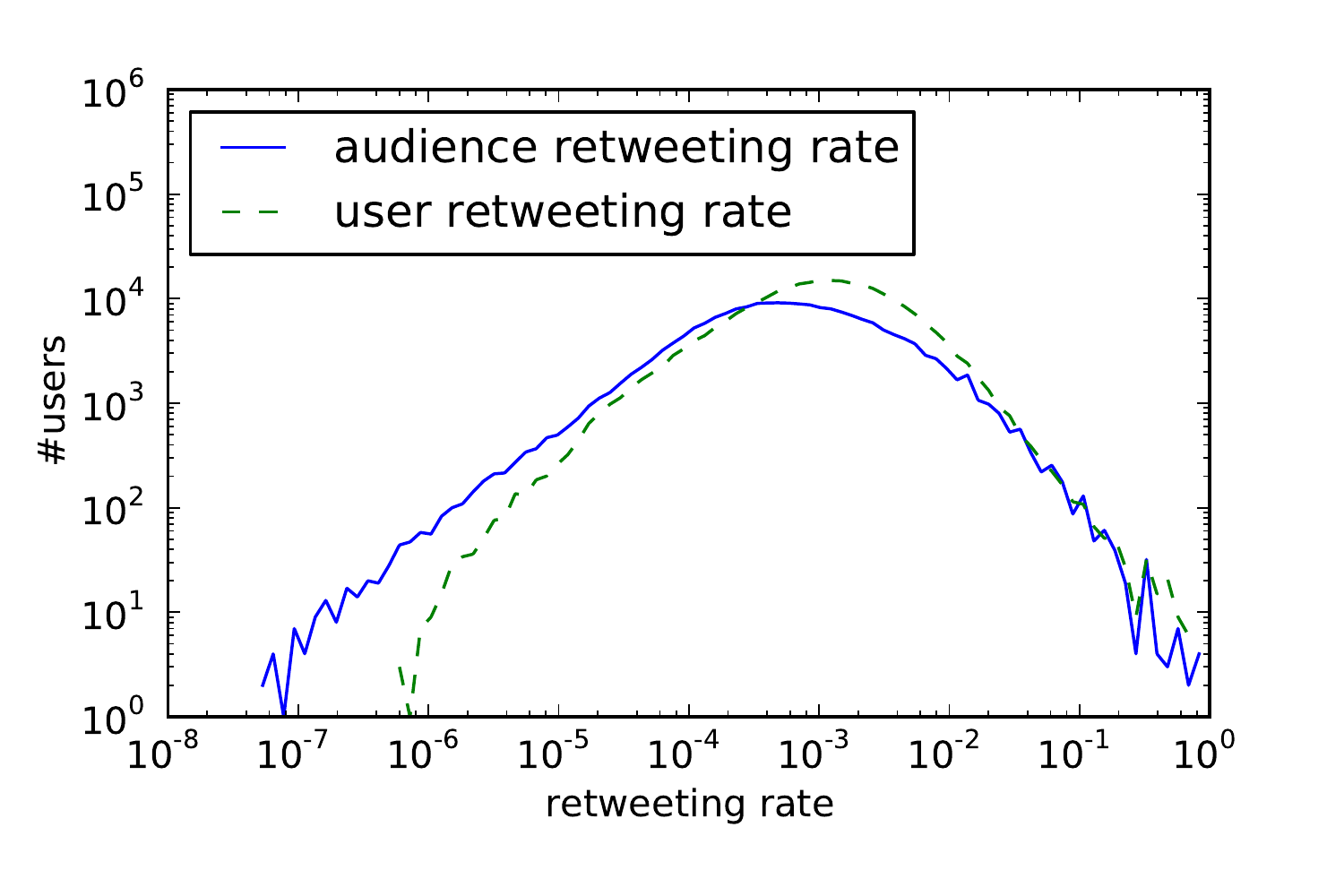}
    \caption{\textbf{Evidence for the Twitter user passivity.}
    \footnotesize{We measure passivity by two metrics: 1. the user retweeting
    rate and 2. the audience retweeting rate. The \emph{user retweeting rate} is
    the ratio between the number of URLs that user $i$ decides to retweet to the
    total number of URLs user $i$ received from the followed users. The \emph{audience retweeting rate} is the
    ratio between the number of user $i$'s URLs that were retweeted by $i$'s followers
    to the number of times a follower of $i$ received a URL from $i$. }}
    %\vspace{-12pt}
\end{figure}

\textbf{Evidence for passivity.} The users that receive information from other
users may never see it or choose to ignore it. We have quantified the degree to
which this occurs on Twitter (Fig. \ref{fig-retweet-rate}). An average Twitter
user retweets only one in 318 URLs, which is a relatively low value. The
retweeting rates vary widely across the users and the small number of the most
active users play an important role in spreading the information in Twitter.
This suggests that the level of user passivity should be taken into
account for the information spread models to be accurate.

\textbf{Assumptions.} Twitter is used by many people as a tool for spreading
their ideas, knowledge, or opinions to others. An interesting and important
question is whether it is possible to identify those users who are very good at
spreading their content, not only to those who choose to follow them, but to a
larger part of the network. It is often fairly easy to obtain information about
the pairwise influence relationships between users. In Twitter, for example,  one
can measure how much influence user A has on user B by counting the number of
times B retweeted A. However, it is not very clear how to use the pairwise
influence information to accurately obtain information about the relative
influence each user has on the whole network. To answer this question we design
an algorithm (IP) that assigns a relative \emph{influence} score and a
\emph{passivity} score to every user. The passivity of a user is a measure of how
difficult it is for other users to influence him. We assume that the influence of
a user depends on both the quantity and the quality of the audience she
influences. In general, our model makes the following assumptions:
\begin{enumerate}

\item A user's influence score depends on the number of people she influences as
well as their passivity.

\item A user's influence score depends on how dedicated the people she
influences are. Dedication is measured by the amount of attention a user pays to a given one
as compared to everyone else.

\item A user's passivity score depends on the influence of those who she's
exposed to but not influenced by.

\item A user's passivity score depends on how much she rejects other user's
influence compared to everyone else.

\end{enumerate}

\textbf{Operation.} The algorithm iteratively computes both the passivity and
influence scores simultaneously in a similar manner as in the HITS algorithm for finding
authoritative web pages and hubs that link to them \cite{HITS}.

Given a weighted directed graph $G =(N,E,W)$ with nodes $N$, arcs $E$, and arc
weights $W$, where the weights $w_{ij}$ on arc $e=(i,j)$ represent the ratio of
influence that $i$ exerts on $j$ to the total influence that $i$ attempted to
exert on $j$, the IP algorithm outputs a function $I:N \rightarrow [0,1]$, which
represents the nodes' relative influence on the network, and a function $P:N
\rightarrow [0,1]$ which represents the nodes' relative passivity of the network.

For every arc $e=(i,j) \in E$, we define the \emph{acceptance rate} by $u_{ij} =
\displaystyle \frac{w_{i,j}}{\displaystyle \sum_{k:(k,j)\in E}w_{kj}}$. This
value represents the amount of influence that user $j$ accepted from  user $i$
normalized by the total influence accepted by $j$ from all users on the network.
The acceptance rate can be viewed as the dedication or loyalty user $j$ has to user $i$. On
the other hand, for every $e=(j,i) \in E$ we define the \emph{rejection rate} by
$v_{ji} = \displaystyle \frac{1-w_{ji}}{\ \displaystyle \sum_{k:(j,k)\in
E}(1-w_{jk})}$. Since the value $1-w_{ji}$ is amount of influence that user $i$
rejected from $j$, then the value $v_{ji}$ represents the influence that user $i$
rejected from user $j$ normalized by the total influence rejected from $j$ by all
users in the network.

The algorithm is based on the following operations: 
\begin{eqnarray}
I_i & \leftarrow & \displaystyle \sum_{j:(i,j)\in E}u_{ij}P_j \\
P_i & \leftarrow & \displaystyle \sum_{j:(j,i)\in E}v_{ji}I_j
\end{eqnarray}

Each term on the right hand side of the above operations corresponds to one of
the listed assumptions. In operation 1 the term $P_j$ corresponds to assumption 1
and the term $u_{ij}$ corresponds to assumption 2. In operation 2 the term $I_j$
corresponds to assumption 3 and the term $v_{ji}$ corresponds to assumption 4.
The \emph{Influence-Passivity algorithm} (Algorithm \ref{ip-algo}) takes the 
graph $G$ as the input and computes the influence and passivity for
each node in $m$ iterations.

\begin{algorithm}[t] 
\caption{The Influence-Passivity (IP) algorithm}
\label{ip-algo}
$I_0 \leftarrow (1,1,\dots,1) \in \mathbf{R}^{|N|}$\;
$P_0 \leftarrow (1,1,\dots,1) \in \mathbf{R}^{|N|}$\;
\For{$i=1$ to $m$}{
Update $R_i$ using operation $(2)$ and the values $I_{i-1}$\;
Update $I_i$ using operation $(1)$ and the values $R_{i}$\;	
\For{$j=1$ to $|N|$}{
	$I_j = \displaystyle \frac{I_j}{\displaystyle \sum_{k \in N}I_k}$\;
	$P_j = \displaystyle \frac{R_j}{\displaystyle \sum_{k \in N}R_k}$\;
}
}
Return $(I_m,P_m)$\;
\end{algorithm}	

\textbf{Generating the input graph.} There are many ways of defining the
influence graph $G=(N,E,W)$. We choose to construct it by taking into account retweets and the
follower graph in the following way: The nodes are users who tweeted at least 3
URLs. The arc $(i,j)$ exists if user $j$ retweeted a URL posted by user $i$ at
least once.  The arc $e=(i,j)$ has weight $w_e = \frac{S_{ij}}{Q_{ij}}$ where
$Q_{i}$ is the number of URLs that $i$ mentioned and $S_{ij}$ is the number of
URLs mentioned by $i$ and retweeted by $j$.

\section{Evaluation} \label{sec-evaluation}

\subsection{Computations}
\begin{figure}
   \centering
     \includegraphics[width=0.46\textwidth]{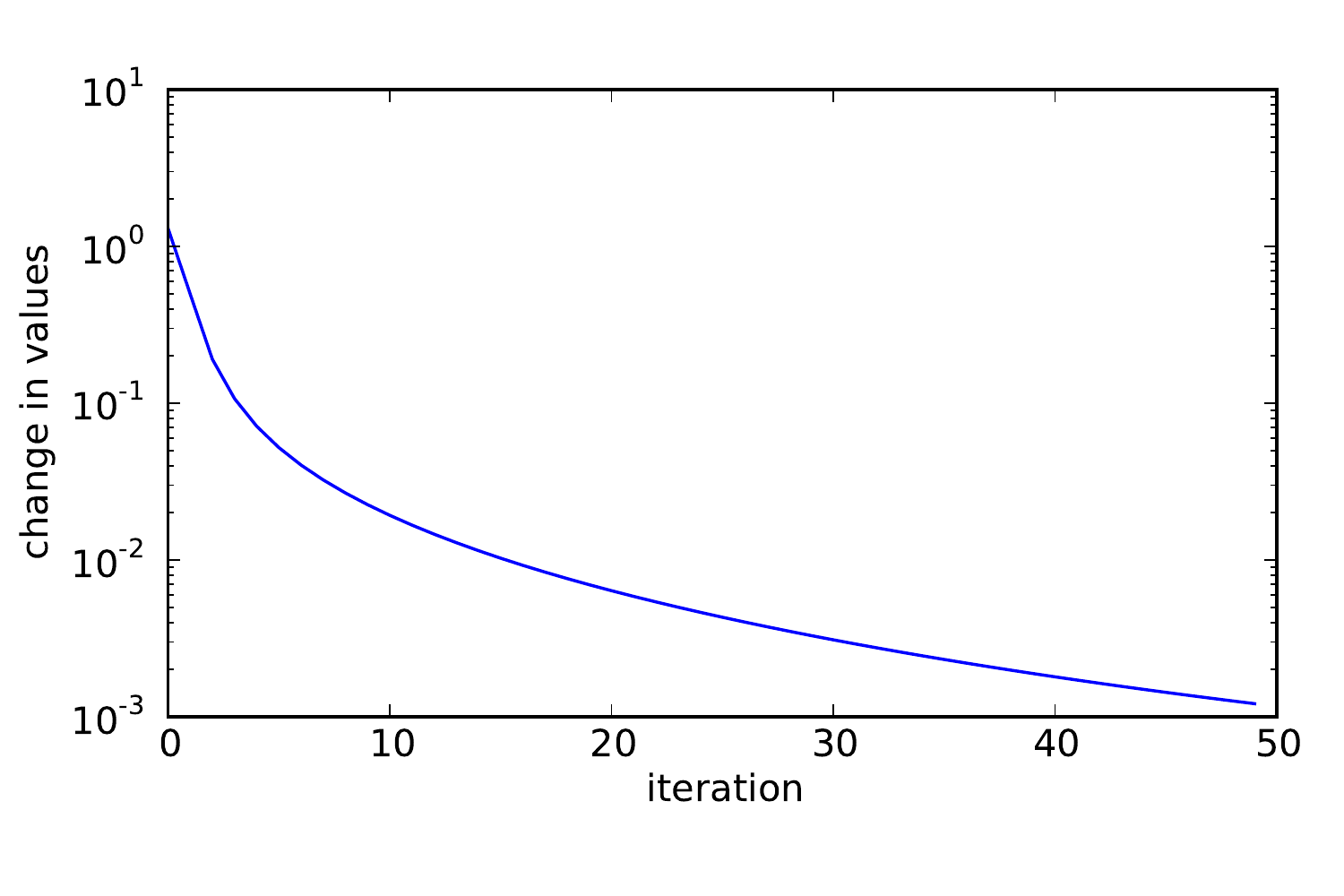}
    \caption{\textbf{IP-algorithm convergence.} \footnotesize{In each iteration
    we measure the sum of all the absolute changes of the computed influence and passivity values since
    the previous iteration}}
    \label{fig-ir-convergence}
  %\vspace{-12pt}
\end{figure}

Based on the obtained dataset (\S\ref{sec-twitter-dataset}) we generate the
weighted graph using the method described in \S\ref{sec-algorithm}. The graph
consists of approximately 450k nodes and 1 million arcs with mean weight of 0.07, and we use it
to compute the PageRank, influence and passivity values for each
node. The Influence-Passivity algorithm (Algorithm \S\ref{ip-algo}) converges to
the final values in tens of iterations (Fig. \ref{fig-ir-convergence}).

\textbf{PageRank.} The PageRank algorithm has been widely used to rank web pages as well as people
based on their authority and their influence \cite{PageRank, Weng2010}. In order to compare
it with the results from the IP algorithm, we compute PageRank on the weighted
graph $G = (N,E,W)$ with a small change. First, since the arcs $e = (i,j) \in E$
indicate that user $i$ exerts some influence on user $j$ then we invert all the
arcs before running PageRank on the graph while leaving the weights intact. In
other words, we generate a new graph $G'=(N',E',W')$ where $N'=N$, $E' =
\{(i,j):(j,i) \in E\}$, and for each $(i,j) \in E'$ we define $w'_{ij} = w_{ji}$.
 This generates a new graph $G'$ analogous to $G$ but where the influenced users
point to their influencers. Second, since the graph $G'$ is weighted we assume
that when the the random surfer of the PageRank algorithm is currently at the
node $i$, she chooses to visit node $j$ next with probability $\displaystyle \frac{w'_{ij}}{\displaystyle \sum_{k:(i,k) \in E'}w'_{ik}}$.

\textbf{The Hirsch Index.} The Hirsch index (or H-index) is used in the
scientific community in order to measure the productivity and impact of a
scientist. A scientist has index $h$ if he has published $h$ articles which
have been cited at least $h$ times each. It has been shown that the H-index is a good
indicator of whether a scientist has had high achievements such as getting the
Nobel prize \cite{Hirsch}. Analogously, in Twitter, a user has index $h$ if $h$
of his URL posts have been retweeted at least $h$ times each.

\subsection{Influence as a correlate of attention}
\label{sec-eval-influence-vs-attention}

\begin{figure*}
   \centering
     \subfigure[Average number of followers vs. number of clicks on URLs]{
       \label{fig-indeg-clicks}
          \includegraphics[width=0.46\textwidth]{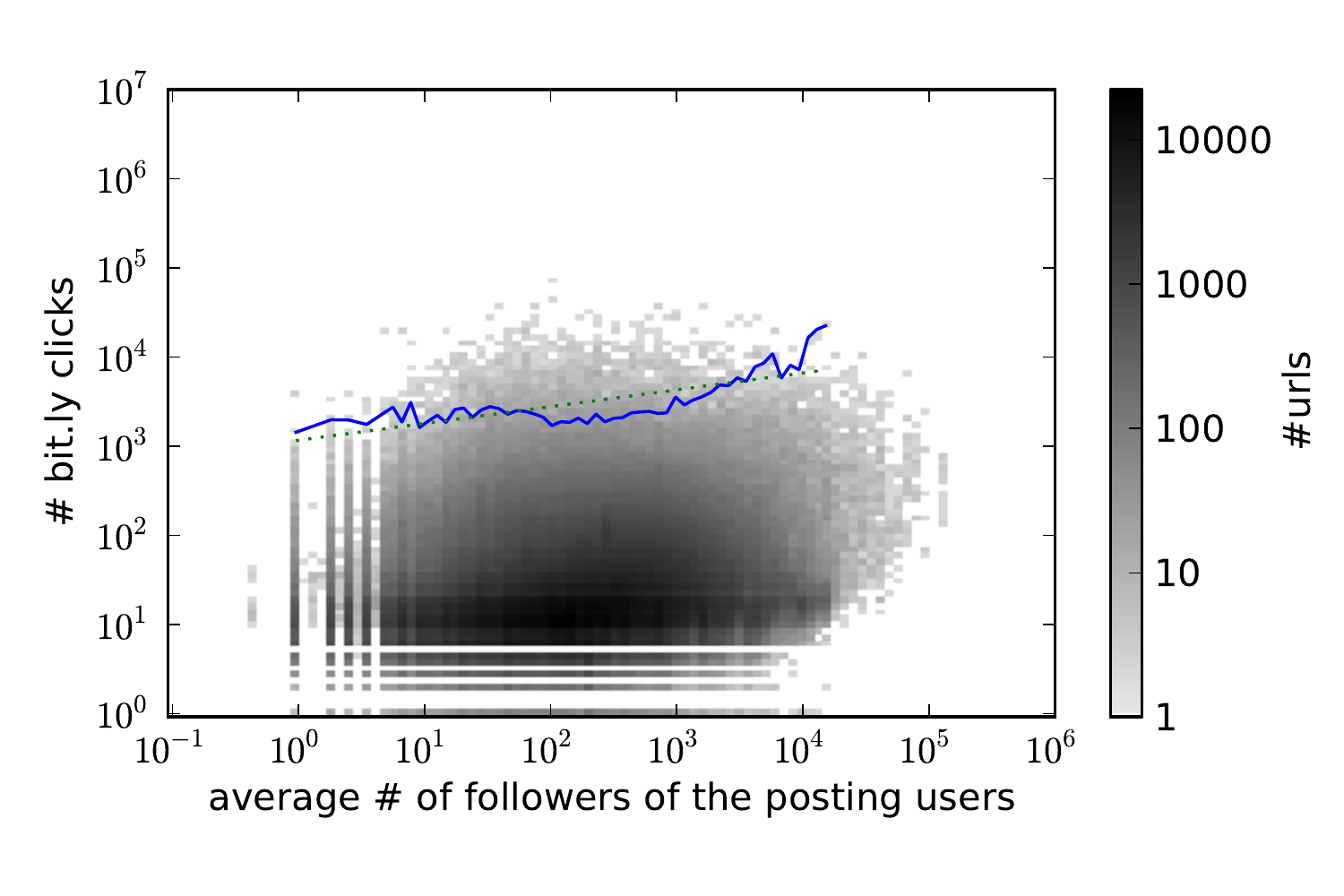}}
    \subfigure[Average number of times users were retweeted vs. number of clicks
    on URLs]{
       \label{fig-retweets-clicks}
    \includegraphics[width=0.46\textwidth]{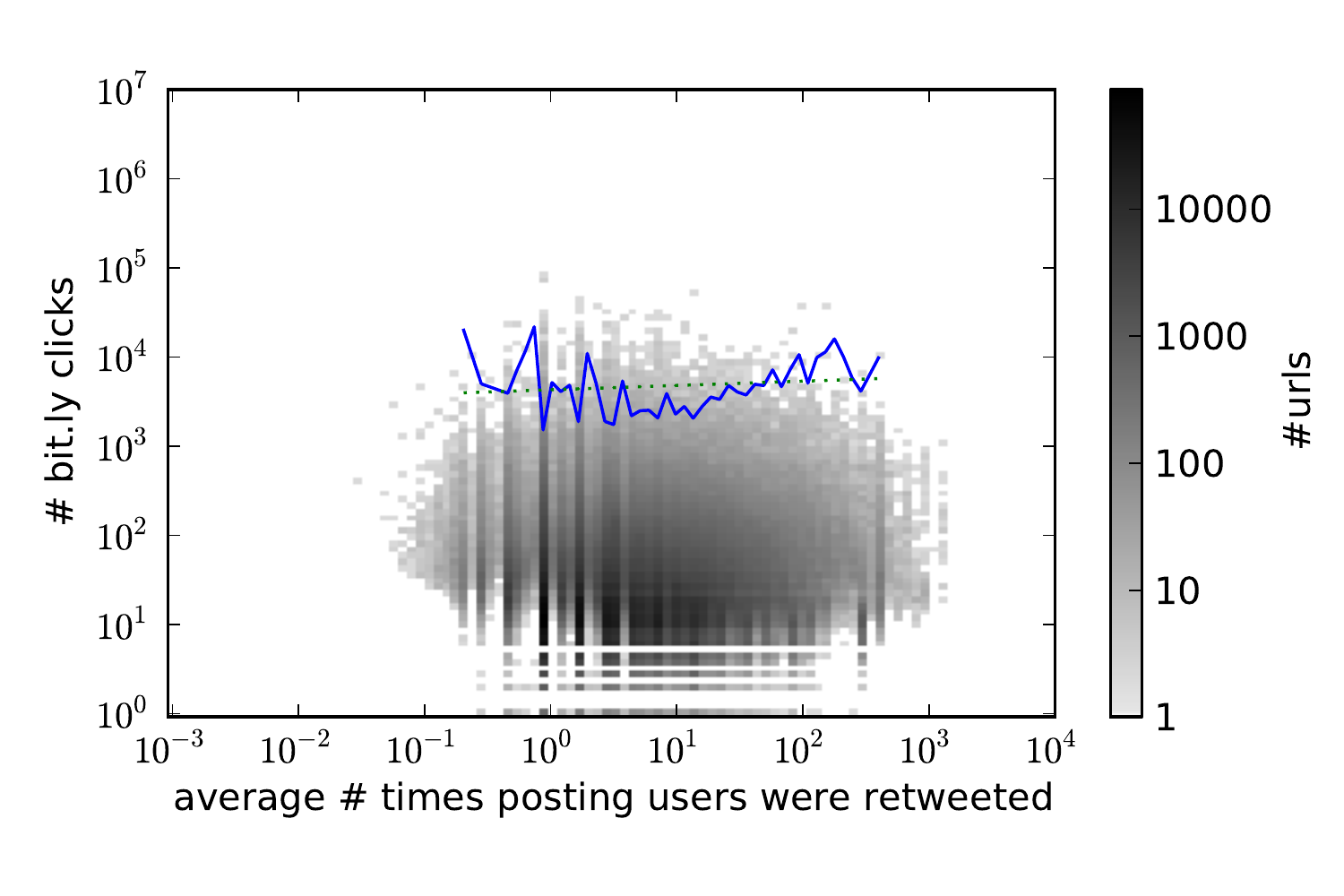}} 
    \subfigure[Average user PageRank vs. number of clicks
    on URLs]{ 
       \label{fig-pagerank-clicks}
    \includegraphics[width=0.46\textwidth]{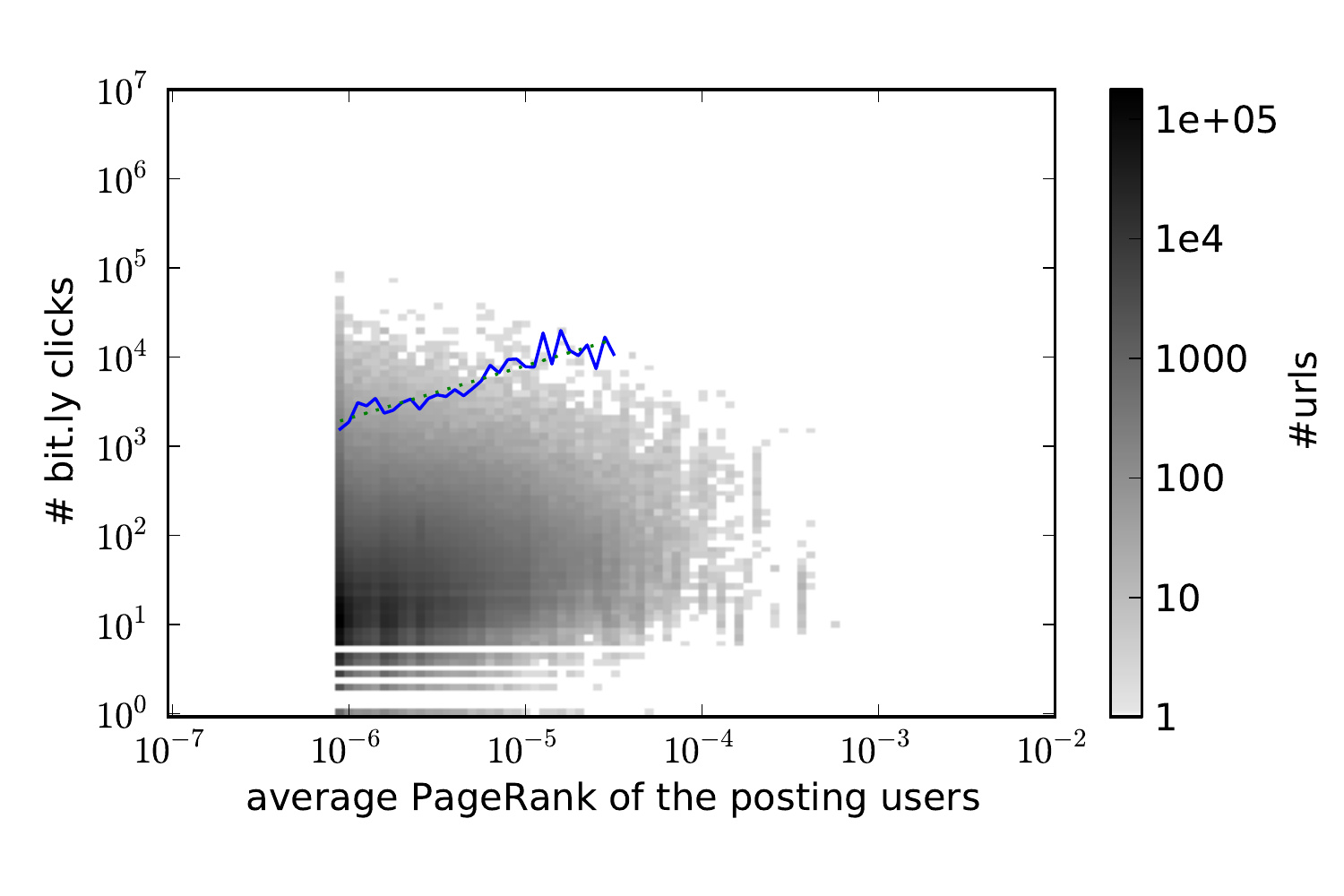}}
     \subfigure[Average user H-index vs. number of clicks on URLs]{   
       \label{fig-hindex-clicks}
    \includegraphics[width=0.46\textwidth]{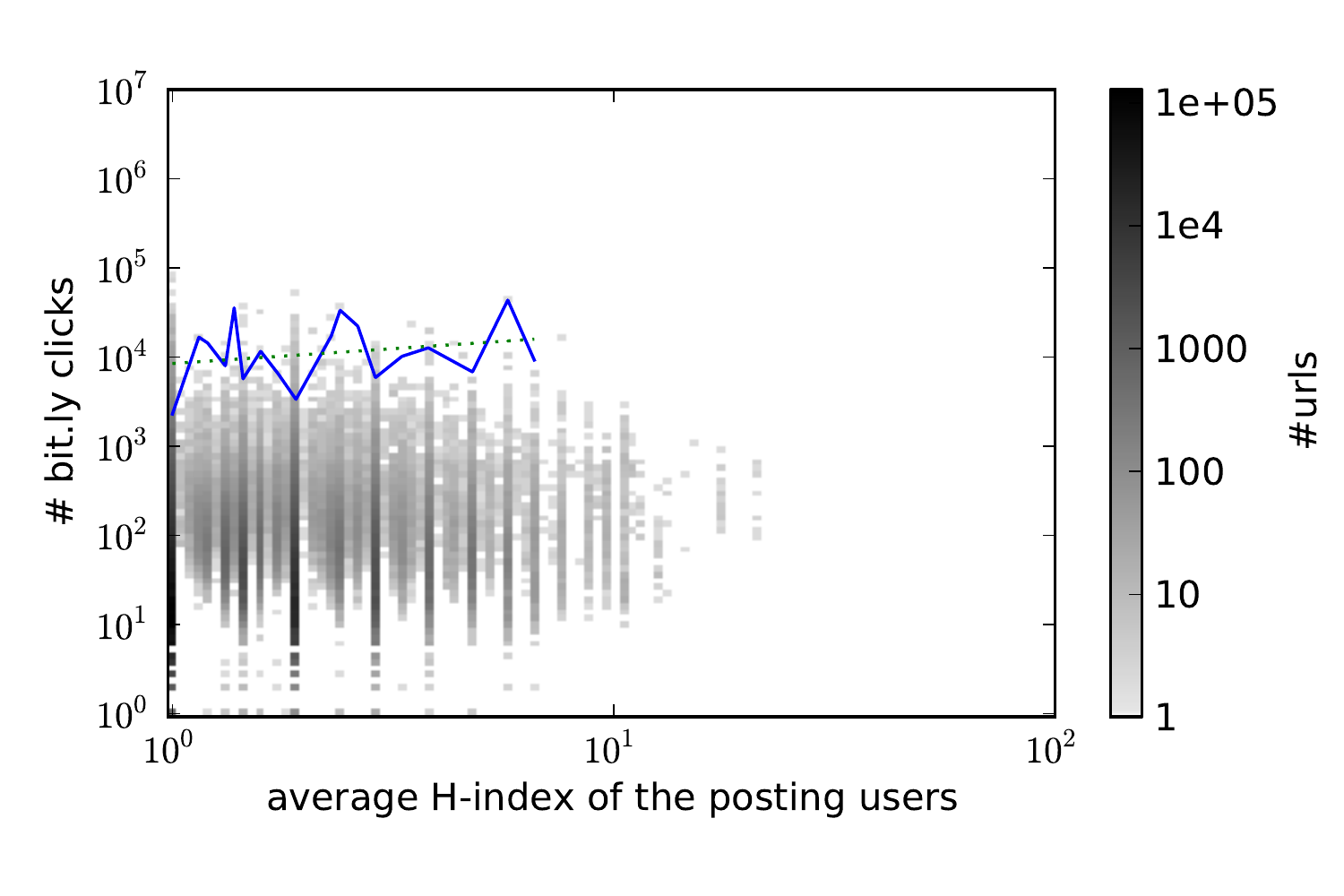}}
    \subfigure[Average user IP-influence vs. number of clicks on URLs, using
    the retweet graph as input]{
       \label{fig-ir-influence-clicks}
    \includegraphics[width=0.46\textwidth]{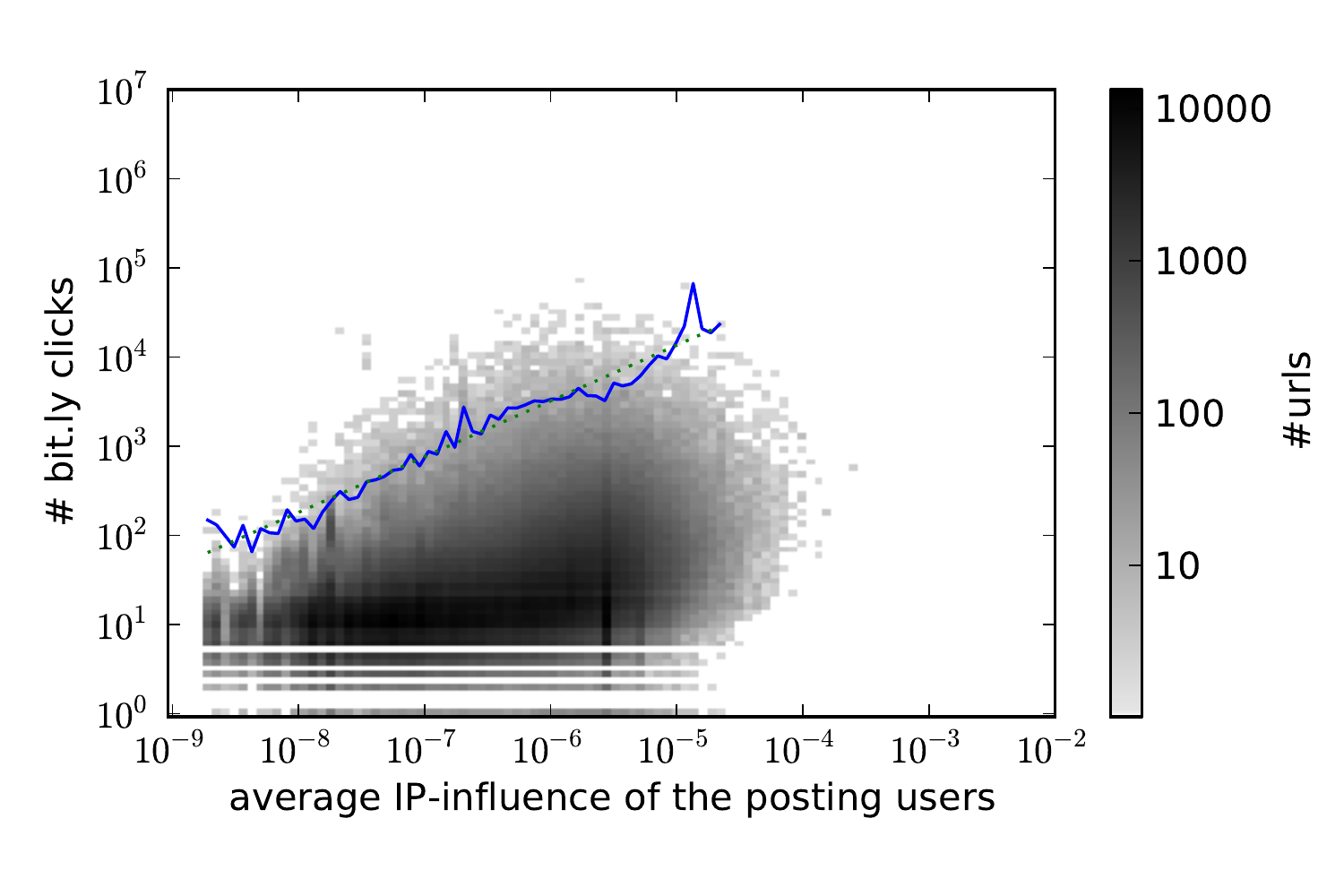}}%\vspace{-5pt}
    \subfigure[Average user IP-influence vs. number of clicks on URLs, using
    the co-mention graph as input]{
       \label{fig-ir-influence-nonrt-clicks}
    \includegraphics[width=0.46\textwidth]{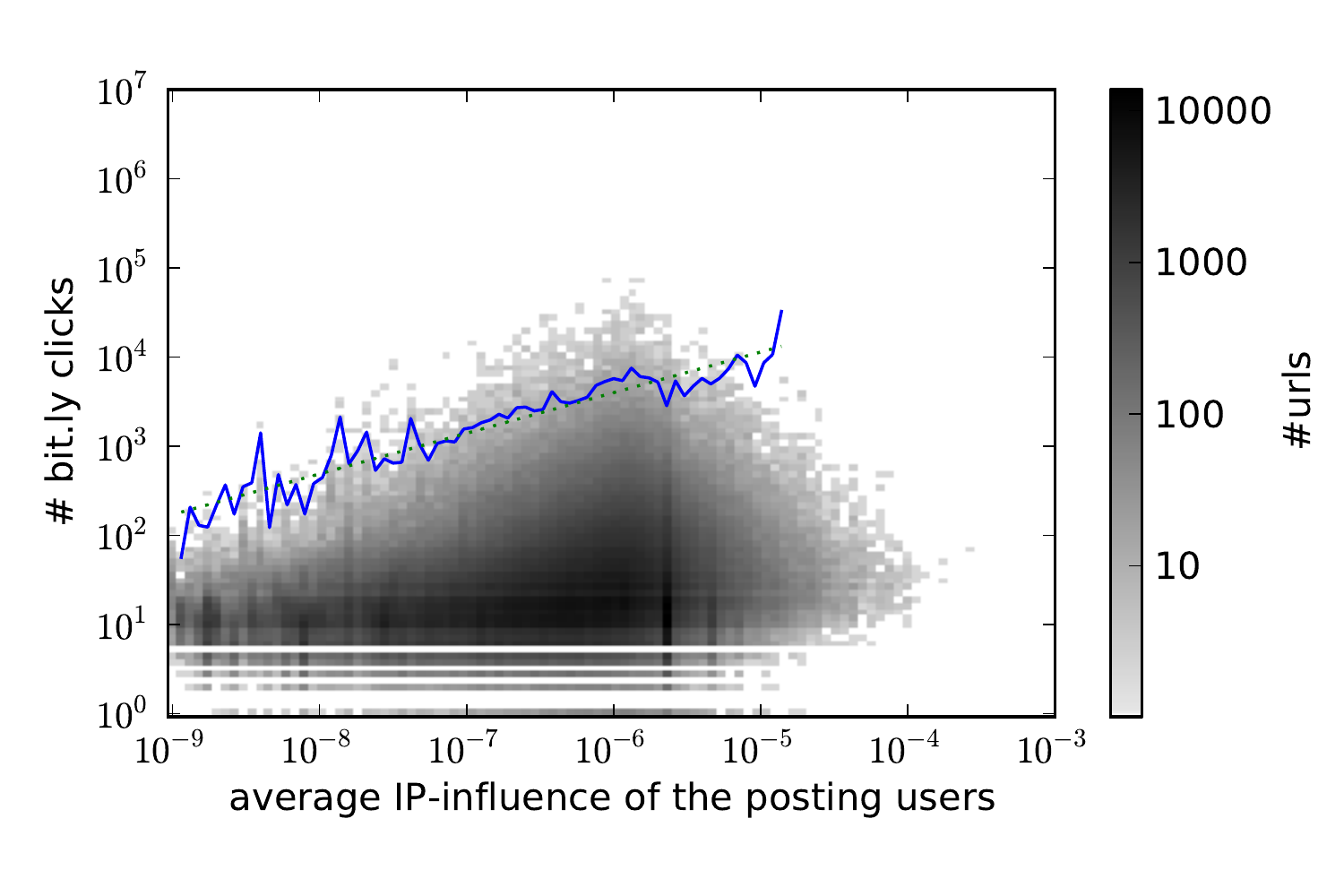}}
    
    \caption{\footnotesize{We consider several user attributes: the number of
    followers, the number of times a user has been retweeted, the user's
    PageRank, H-index and IP-influence. For each of the 3.2M Bit.ly URLs we
    compute the average value of a user's attribute among all the users that
   mentioned that URL. This value becomes the $x$ coordinate of the URL-point;
   the $y$ coordinate is the number of clicks on the Bit.ly URL. The
   density of the URL-points is then plotted for each of the four user attributes.
   The solid line in each figure represents the $99.9th$ percentile of
   Bit.ly clicks at a given attribute value. The dotted line is the
   linear regression fit for the solid line.}} \label{fig-influence-clicks}
  %\vspace{-12pt}
\end{figure*}

Any measure of influence is necessarily a subjective one. However, in this case,
a good measure of influence should have a high predictive power on how well the
URLs mentioned by the influential users attract attention and propagate in the
social network. We would expect the URLs that highly influential users propagate
to attract a lot of attention and user clicks. Thus, a viable estimator of
attention is the number of times a URL has been accessed.

\textbf{Click data.} Bit.ly is a URL shortening service that for each shortened
URL keeps track of how many times it has been accessed. For the 3.2M
Bit.ly URLs found in the tweets we have queried the Bit.ly API for the number of
clicks the service has registered on that URL. 

\textbf{URL traffic correlation.} Using the URL click data, we take several
different user attributes and test how well they can predict the attention the
URLs posted by the users receive (Fig. \ref{fig-influence-clicks}). It is
important to note that none of the influence measures are capable of predicting
the exact number of clicks. The main reason for this is that the amount of
attention a URL gets is not only a function of the influence of the users
mentioning it, but also of many other factors including the virality of the URL
itself and more importantly, whether the URL was mentioned anywhere outside of
Twitter, which is likely to be the biggest source of unpredictability in the
click data. The click data that we collected represents the total clicks on the
URLs.

The wide range of factors potentially affecting the 
Bit.ly clicks may prevent us from predicting their number accurately. However,
the upper bound on that number can to a large degree be predicted. To eliminate the
outlier cases, we examined how the $99.9th$ percentile of the clicks varied as
the measure of influence increased. 

\textbf{Number of followers.} The most readily available and often used by the
Twitterers measure of influence is the number of followers a user has. As the
Figure \ref{fig-indeg-clicks} shows, the number of followers of an average
poster of a given URL is a relatively weak predictor of the maximum number of
clicks that the URL can receive.

\textbf{Number of retweets.} When users post URLs their posts might be retweeted
by other users. Each retweet explicitly credits the
original poster of the URL (or the user from whom the retweeting user heard
about the URL). The number of times a user has been credited in a retweet has been assumed to be a  good measure of influence \cite{Cha2010}. However, Figure \ref{fig-retweets-clicks} shows that
the number of times a user has been retweeted in the past is a poor predictor of
the maximum number of clicks the URLs posted by that user can get.

\textbf{PageRank.} Figure \ref{fig-pagerank-clicks} shows that the average
PageRank of those who tweet a certain URL does not correlate well with the number
of clicks the URL will get. One of the main differences between the IP algorithm
and PageRank is that the IP algorithm takes into account the passivity of the
people a user influences and PageRank does not.IP-influence is a much better
indicator of URL popularity than PageRank. This suggests that influencing users
who are difficult to influence, as opposed to simply influencing many users, has
a positive impact on the eventual popularity of the message that a user tweets.

\textbf{The Hirsch Index.} Figure \ref{fig-hindex-clicks} shows that despite the
fact that in the scientific community the H-index is used as a good predictor of
high achievements, in Twitter it does not correlate well with URL popularity.
This may reflect the fact that attention in the scientific community plays a
symmetric role, since those who pay attention to the work of others also seek it
from the same community. Thus, citations play a strategic role in the successful
publishing of papers, since the expectation of authors is that referees and
authors will demand attention to their work and those of their colleagues. Within
Social Media such symmetry does not exist and thus the decision to forward a
message to the network lacks this particularly strategic value.

\textbf{IP-Influence score.} As we can see in Figure
\ref{fig-ir-influence-clicks}, the average IP-influence of those who tweeted a
certain URL can determine the maximum number of clicks that a URL will get. Since
the URL clicks are never considered by the IP algorithm to compute the user's
influence, the fact that we find a very clear connection between average
IP-influence and the eventual popularity of the URLs (measured by clicks) serves as 
an unbiased evaluation of the algorithm and exposes the power of IP-influence.
For example, as we can see in Figure  \ref{fig-ir-influence-clicks}, given a
group of users having very large average IP-influence scores who post a URL we can
estimate, with $99.9\%$ certainty, that this URL will not receive more than
$100,000$ clicks. On the other hand, if a group of users with very low average
IP-influence score post the same URL we can estimate, with $99.9\%$
certainty that the URL will not receive more than $100$ clicks.

\begin{figure}[t]
   \centering
    \includegraphics[width=0.46\textwidth]{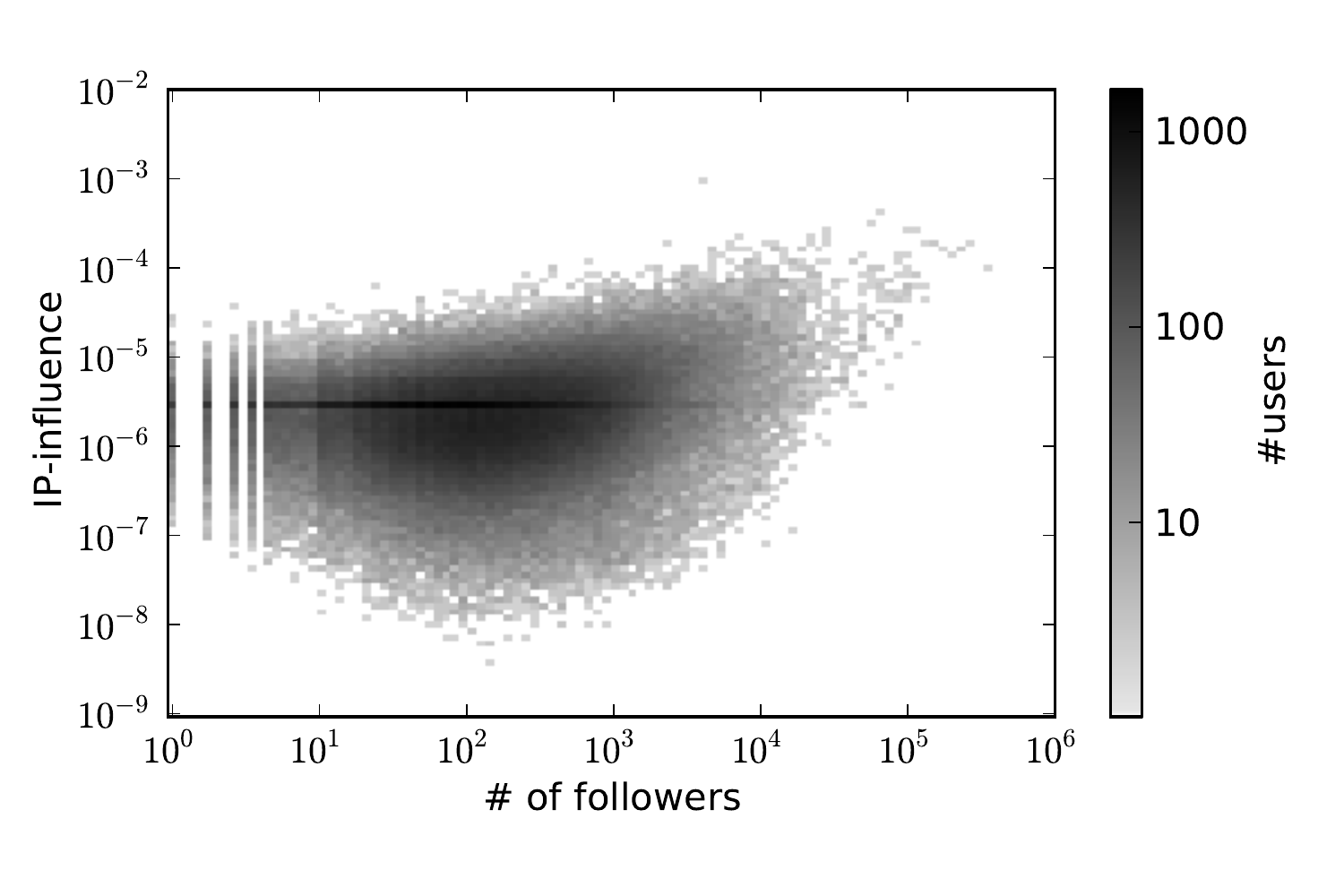}
    \caption{\footnotesize{For each user we place a user-point with
    IP-influence as the $y$ coordinate and the $x$ coordinate set to the
    number of user's followers. The density of user-points is
    represented in grayscale. }}
    \label{fig-indeg-ir}
  %\vspace{-12pt}
\end{figure}

Furthermore, figure \ref{fig-indeg-ir} shows that a user's IP-influence is not
well correlated with the number of followers she has. This reveals interesting
implications about the relationship between a person's popularity and the
influence she has on other people. In particular, it shows that having many
followers on Twitter does not imply power to influence them to even
click on a URL.

\section{IP Algorithm Adaptability} \label{sec-robustness}

\begin{figure}[t]
   \centering
          \includegraphics[width=0.46\textwidth]{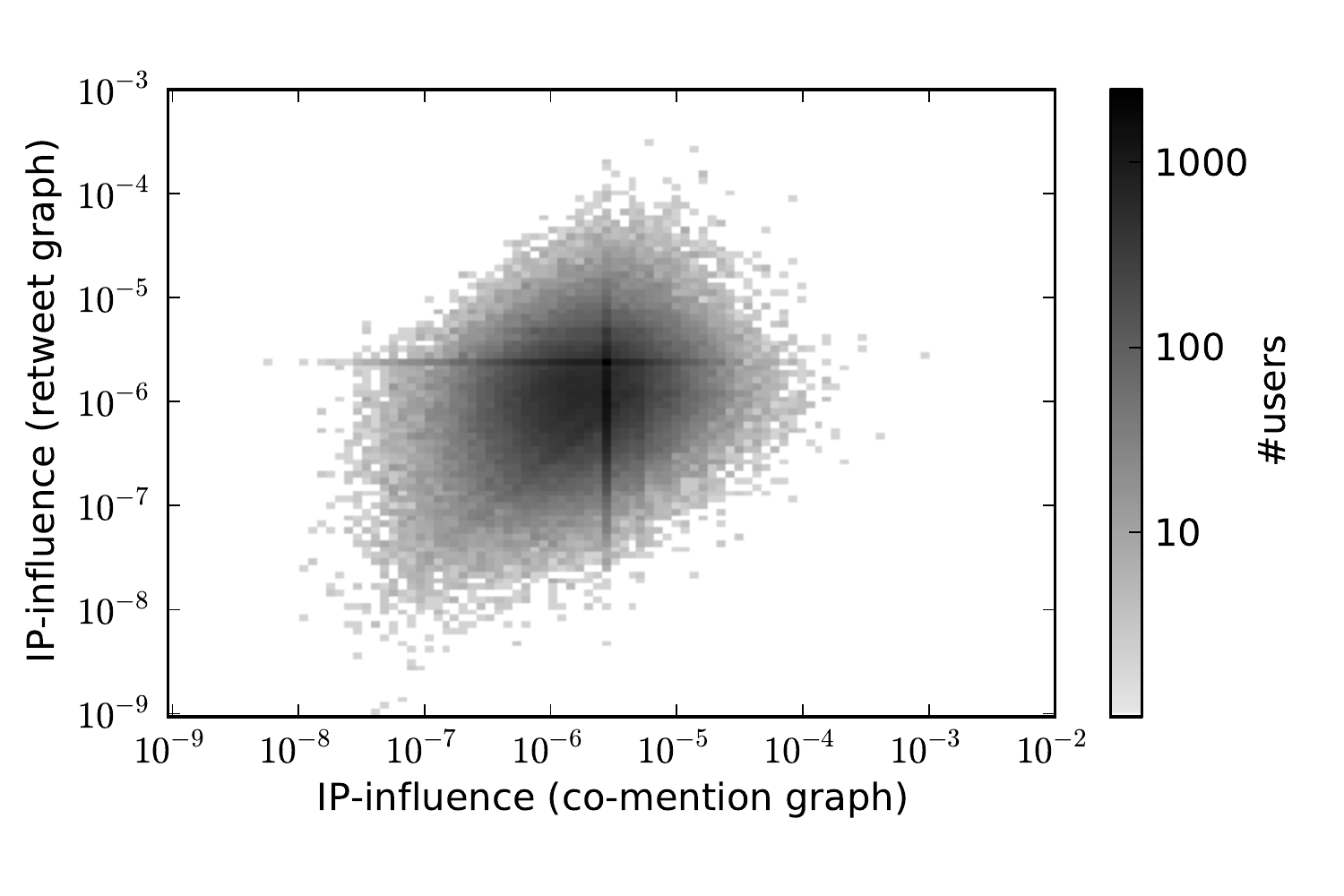}
    %vspace{-5pt}
    \caption{\footnotesize{The correlation between the IP-influence values
    computed based on two inputs: the co-mention influence graph and the
    retweet influence graph.}}
    \label{fig-nonrt-rt-correlation}
  %\vspace{-12pt}
\end{figure}

As mentioned earlier (\S\ref{sec-algorithm}) there are many ways of defining a
social graph in which the edges indicate pairwise influence. We have so far been
using the graph based on which user retweeted which user (\emph{retweet influence
graph}). However, that explicitly signals of influence such as retweets are not
always available.  One way of overcoming this obstacle is to use other, possibly
weaker, signals of influence. In the case of Twitter, we can define an influence
graph based on mentions of URLs without regard of actual retweeting in the
following way.

\textbf{The co-mention graph.} The nodes of the \emph{co-mention influence
graph} are users who tweeted at least three URLs. The edge $(i,j)$ exists if user $j$ follows
user $i$ and $j$ mentioned at least one URL that $i$ had previously mentioned.  The edge
$e=(i,j)$ has weight $w_e = \frac{S_{ij}}{F_{ij}+S_{ij}}$ where $F_{ij}$ is the
number of URLs that $i$ mentioned and $j$ never did and $S$ is the number of URLs
mentioned by $j$ and previously mentioned by $i$.

The resulting graph has the disadvantage that the edges are based on a much less
explicit notion of influence than when based on retweets. Therefore the
graph could have edges between users who do not influence each other.  On the
other hand, the retweeting conventions on Twitter are not uniform and therefore
sometimes users who repost a URL do not necessarily credit the correct source of
the URL with a retweet \cite{retweet}. Hence, the influence graph based on
retweets  has potentially missing edges.

Since the IP algorithm has the flexibility of allowing any influence graph as
input, we can compute the influence scores  of the users based on the
co-mention influence graph and compare with the results obtained from the
retweet influence graph. As we can see in Figure
\ref{fig-ir-influence-nonrt-clicks}, we find that the retweet graph yields influence scores that are better at
predicting the maximum number of clicks a URL will obtain than the
co-mention influence graph. Nevertheless, Figures
\ref{fig-ir-influence-nonrt-clicks}, \ref{fig-pagerank-clicks}, \ref{fig-indeg-clicks},
and \ref{fig-retweets-clicks} show that the influence values obtained from the
co-mention influence graph are still better at predicting URL traffic than
other measures such as PageRank, number of followers, H-index or the total number of
times a user has been retweeted. Furthermore, Figure
\ref{fig-nonrt-rt-correlation} shows that the influence score based on both graphs  do not correlate well, which
suggests that considering explicit vs. implicit signals of influence can
drastically change the outcome of the IP algorithm. In general, we find that the
explicitness of the signal provided by the retweets yields slightly better
results when it comes to predicting URL traffic, however, the influence
scores based on co-mentions may surface a different set of potentially
influential users.

\section{Case Studies} \label{caseStudy}
As we mentioned earlier, one important application of the IP algorithm is
ranking users by their relative influence. In this section, we present a series of
rankings of Twitter users based on the influence, passivity, and number of
followers.

\textbf{The most influential.} Table 1 shows the users with the most
IP-influence in the network. We constrain the number of URLs posted to 10 to
obtain this list, which is dominated by news services from politics, technology,
and Social Media. These users post many links which are forwarded by other
users, causing their influence to be high.
\begin{table}%[htdp]
	\begin{center}
	\begin{tabular}{|l|l|} 
		\hline	
		mashable & Social Media Blogger \\
		jokoanwar & Film Director \\
		google & Google News \\
		aplusk & Actor\\
		syfy & Science Fiction Channel  \\
		smashingmag & Online Developer Magazine \\
		michellemalkin & Conservative Commentator \\
		theonion & News Satire Organization \\
		rww & Tech/Social Media Blogger \\
		breakingnews & News Aggregator \\
		\hline
		\end{tabular}
		\caption{Users with the most IP-influence (with at least 10 URLs posted in the
		period)}
		\label{label1}
	\end{center}
\end{table}

\textbf{The most passive.} Table 2 shows the users with the most
IP-passivity in the network. Passive users are those who follow many people,
but retweet a very small percentage of the information they consume.
Interestingly, robot accounts (which automatically aggregate keywords or
specific content from any user on the network), suspended accounts (which are
likely to be spammers), and users who post extremely often are among the users with the most
IP-passivity. Since robots "attend" to all existing tweets and only retweet
certain ones, the percentage of information they forward from other users is
actually very small. This explains why the IP-algorithm assigns them such high
passivity scores. This also highlights a new application of the IP-algorithm: 
automatic identification of robot users including aggregators and
spammers. 

\begin{table}[htdp]
	\begin{center}
		\begin{tabular}{|l|l|}
			\hline 
			redscarebot & Keyword Aggregator \\			drunk\_bot & Suspended \\ 
			tea\_robot & Keyword Aggregator \\ 
			condos & Listing Aggregator \\
			wootboot & Suspended \\
			raybeckerman & Attorney \\ 
			hashphotography & Keyword Aggregator \\			charlieandsandy & Suspended \\ 
			ms\_defy & Suspended \\ 
			rpattinsonbot & Keyword Aggregator \\
			\hline
		\end{tabular} 
		\caption{Users with the most IP-passivity}
	\end{center}
	\label{MostPass}
\end{table}

\textbf{The least influential with many followers.} We have demonstrated that the
amount of attention a person gets may not be a good indicator of the influence
they have in spreading their message. In order to make this point more explicit,
we show, in Table 3, some examples of users who are followed by
many people but have relatively low influence. These users are very popular and
have the attention of millions of people but are not able to spread their message
very far. In most cases, their messages are consumed by their followers but not
considered important enough to forward to others.

\begin{table*}[htdp]
	\begin{center}
		\begin{tabular}{|l|l|c|c|}
			\hline
			User name & Category & Rank by \# followers & Rank by IP-influence \\
			\hline 
			thatkevinsmith & Screen Writer & 33 & 1000  \\
			nprpolitics & Political News & 41 & 525\\
			eonline & TV Channel & 42 & 1008 \\
			marthastewart & Television Host & 43 & 1169 \\
			nba & Sports & 64 & 1041\\
			davidgregory & Journalist & 106 &3630  \\
			nfl & Sports & 110 &2244\\ 
			cbsnews & News Channel  & 114 & 2278\\
			jdickerson & Journalist & 147 & 4408 \\
			newsweek & News Magazine & 148 &756\\
			\hline 
		\end{tabular}
		\caption{Users with many followers and low relative influence}
	\end{center}
	\label{HighFolLowInf}
\end{table*}

\textbf{The most influential with few followers.} We are also able identify users
with very low number of followers but high influence. Table 4 shows the users
with the most influence who rank less that $100,000$ in number of followers. We
find that during the data collection period some of the users in this category
ran very successful retweeting contests where users who retweeted their URLs
would have the chance of winning a prize. Moreover, there is a group of users who
post from \emph{Twitdraw.com}, a website where people can make drawings and post
them on Twitter. Even though these users don't have many followers, their
drawings are of very high quality and spread throughout Twitter reaching many
people. Other interesting users such as local politicians and political
cartoonists are also found in the list. The IP-influence measure allows us to
find interesting content posted by users who would otherwise be buried by
popularity rankings such as number of followers.

\begin{table*}[htdp]
	\begin{center}
		\begin{tabular}{|l|l|c|c|}
			\hline
			User name & Category & Rank by \# followers & Rank by IP-influence \\
			\hline 
			cashcycle & Retweet Contest & 153286 & 13 \\
			mobiliens & Retweet Contest & 293455 & 70 \\ 
			jadermattos &  Twitdraw & 227934 & 134\\ 
			\_jaum\_ & Twitdraw  & 404385 & 143 \\ 
			robmillerusmc &  Congressional Candidate & 147803 & 145\\ 
			sitekulite	& Twitdraw & 423917 & 149\\
			jesse\_sublett & Musician & 385265 & 151 \\
			cyberaurora & Tech News Website & 446207 & 163 \\
			viveraxo & Twitdraw & 458279 & 165 \\
			fireflower\_ & Political Cartoons & 452832 & 195\\
			\hline
		\end{tabular}
		\caption{Users with very few followers but high relative
		influence}
	\end{center}
	\label{LowFolHighInf}
\end{table*}

\section{Discussion} \label{sec-label}

\textbf{Influence as predictor of attention.} As we demonstrated in
\S\ref{sec-evaluation}, the IP-influence of the users is an accurate predictor of the
upper bound on the total number of clicks they can get on the URLs they post. The
input to the influence algorithm is a weighted graph, where the arc weights
represent the influence of one user over another. This graph can be derived from
the user activity in many ways, even in cases where explicit feedback in the form
of retweets or ``likes'' is not available (\S\ref{sec-robustness}).

\textbf{Topic-based and group-based influence.} The Influence-Passivity
algorithm can be run on a subpgraph of the full graph or on the subset of the user activity data. For example, if only
users tweeting about a certain topic are part of the graph, the IP-influence
determines the most influential users in that topic. It is an open question
whether the IP algorithm would be equally accurate at different graph scales.

\textbf{Content ranking.} The predictive power of IP-influence can be used for
content filtering and ranking in order to reveal content that is most likely to receive
attention based on which users mentioned that content early on. Similarly, as in the case
of users, this can be computed on a per-topic or per-user-group basis.

\textbf{Content filtering.} We have observed from our passivity experiments that
highly passive users tend to be primarily robots or spammers. This leads to an interesting extension 
of this work to perform content filtering, limiting the tweets to influential users and thereby reducing 
spam in Twitter feeds.

\textbf{Influence dynamics.} We have computed the influence measures over a fixed
300-hour window. However, the Social Media are a rapidly changing, real-time
communication platform. There are several implications of this. First, the IP
algorithm would need to be modified to take into account the tweet timestamps.
Second, the IP-influence itself changes over time, which brings a number of
interesting questions about the dynamics of influence and attention. In
particular, whether users with spikes of IP-influence are overall more
influential than users which can sustain their IP-influence over time is an open
question.

\section{Conclusion} \label{sec-conclusions}

Given the mushrooming popularity of Social Media, vast efforts are devoted by
individuals, governments and enterprises to getting attention to their ideas,
policies, products, and commentary through social networks. But the very large
scale of the networks underlying Social Media makes it hard for any of these
topics to get enough attention in order to rise to the most trending ones. Given
this constraint, there has been a natural shift on the part of the content
generators towards targeting those individuals that are perceived as influential
because of their large number of followers. This study shows that the correlation
between popularity and influence is weaker than it might be expected. This is a
reflection of the fact that for information to propagate in a network,
individuals need to forward it to the other members, thus having to actively
engage rather than passively read it and cease to act on it. Moreover, since our
measure of influence is not specific to Twitter it is applicable to many other
social networks. This opens the possibility of discovering influential
individuals within a network which can on average have a further reach than
others in the same medium, regardless of their popularity.


\bibliographystyle{abbrv}
%\bibliography{sigproc}  % sigproc.bib is the name of the Bibliography in this
% case

% ACM needs 'a single self-contained file'!
%\balancecolumns

%This section is inserted by \LaTeX; you do not insert it.
%You just add the names and information in the
%\texttt{{\char'134}additionalauthors} command at the start
%of the document.
%\subsection{References}
%Generated by bibtex from your ~.bib file.  Run latex,
%then bibtex, then latex twice (to resolve references)
%to create the ~.bbl file.  Insert that ~.bbl file into
%the .tex source file and comment out
%the command \texttt{{\char'134}thebibliography}.

\end{document}